\documentclass[11pt]{article}
 
\usepackage{amsmath}
 \usepackage{tabularx}
 \usepackage{array}
 \usepackage{booktabs}
 \usepackage{graphicx}% Include figure files
 \usepackage{longtable}
 \def\CP{{$\cal CP~$}}
 \usepackage{amsfonts}
 \usepackage{float}
\usepackage{hyperref}
\usepackage{bm, geometry, cite, amssymb}
\begin{document}
 
 \title{\LARGE \bf {Geometric Phases in Two-Level Mixing: From Neutral Mesons to Holonomic Qubits and Topological Majorana Modes}
 }

\author{
Swarup Sangiri$^{1}$\thanks{swarupsangiri@iitkgp.ac.in, swarup.phys@gmail.com},
 Utpal Sarkar$^{2}$,
and A Taraphder$^{1}$
}

\date{}

\maketitle

\begin{center}
\small
$^{1}$ Department of Physics, Indian Institute of Technology Kharagpur, Kharagpur, 721302, India\\
$^{2}$ Department of Physical Sciences, 
Indian Institute of Science Education and Research Kolkata,\\ Mohanpur 741246, India
\end{center}

\vspace{0.3cm}
 %\author{
 	%Swarup Sangiri$^{1}$ %\thanks{swarup.phys@gmail.com},
    %Arghya Taraphder$^{1}$,
   % and Utpal Sarkar$^{2}$
   %}
   
%\date{}
%\maketitle
%\noindent
%$^{1}$\textit{Physics Department, Indian Institute of %Technology Kharagpur, Kharagpur, 721302, India}\\
%\noindent
%$^{2}$\textit{Department of Physical Sciences, 
%Indian Institute of Science Education and Research %Kolkata,\\ Mohanpur 741246, India}

\begin{abstract}
We investigate the geometric structure of a Hermitian two-level mixing
Hamiltonian motivated by neutral meson oscillations and its connections with
qubit, fermionic, and topological descriptions. Mapping the Hamiltonian to
an effective qubit representation, we analyze the geometric phase associated
with cyclic parameter evolution on the Bloch sphere. Without identifying the
complex mixing phase with a physical \CP-violation observable, we interpret
it as a \CP-like geometric parameter controlling the azimuthal orientation
of the Hamiltonian vector and its reversal under phase conjugation.
Third-order Bargmann invariants constructed from the eigenstates yield an
associated discrete phase whose dependence on the mixing parameters is
examined alongside the continuous geometric construction. The cyclic
evolution is also represented as a single-qubit $R_z$ phase operation and
expressed through Majorana bilinears. Extending the construction to the
momentum-dependent Bogoliubov--de Gennes Hamiltonian of the Kitaev chain, we
relate momentum-space winding to the topological regimes and the quantized
Berry (Zak) phase in the real-parameter model, while the corresponding
Bargmann construction approaches the global phase in the continuum limit.
Together, these results provide a geometric perspective connecting phase
structure, qubit operations, fermionic representations, and momentum-space
topology within related two-level Hamiltonians.
\end{abstract}
\newpage

\section{Introduction}
Quantum coherence and interference phenomena play a central role in both particle physics and quantum information science, particularly in systems governed by two-level quantum dynamics. In particle physics, neutral meson systems such as $K^0-\overline{K^0}$,  $B^0-\overline{B^0}$ and $D^0-\overline{D^0}$ exhibit oscillatory behavior arising from the nontrivial structure of the effective Hamiltonian \cite{BigiSanda2009, LeeOehmeYang1957}. The theoretical description of such phenomena was developed within the framework of flavor physics and the study of \CP violation \cite{GellMannPais1955, Christenson1964, Wolfenstein1964, KobayashiMaskawa1973}.

At a formal level, the dynamics of neutral meson mixing can be described by
an effective $2\times2$ Hamiltonian acting on the flavor basis
\cite{Nir1999, Sarkar2008}. The resulting two-level structure provides a
natural setting in which the evolution of the system can be represented
geometrically, with the Hamiltonian parameters determining the orientation
of the corresponding eigenstate rays in a projective two-level space.
Geometric phases can arise from the evolution of quantum states in parameter
space, with cyclic evolution providing a particularly important setting for
their study \cite{Berry1984, Simon1983, Anandan1992}. In the present
work, we study this geometric structure within the Hermitian, decay-free
limit of the mixing Hamiltonian, where the two-level dynamics admit a
direct representation in terms of Bloch-sphere geometry. 

Interestingly, the same two-level structure also underlies the fundamental unit of quantum information, the qubit. For a single qubit, coherent dynamics can be described by a
$2\times2$ Hamiltonian, and the corresponding state evolution can be
visualized on the Bloch sphere. This shared mathematical framework suggests a natural correspondence between two-level systems encountered in particle physics and those used in quantum information theory. In particular, geometric phases acquired during cyclic evolution provide
the basis for geometric phase gates and, more generally, holonomic quantum
computation \cite{ZanardiRasetti1999}.

The Bargmann invariant (BI) provides a discrete characterization of
geometric phase relations associated with closed sequences of quantum rays
\cite{Bargmann1964, Rabei1999}. Its rephasing-invariant formulation has made it a
useful tool for characterizing geometric phases in finite sequences of
quantum states, including in two-level and particle-physics settings
\cite{MukundaSimon1993, SangiriSarkar2023, Sangiri2026, Hetenyi2026}. In the present work, the BI is used to examine the phase structure of
finite state sequences alongside the continuous geometric phase associated
with cyclic evolution. By
constructing the third-order BI for a finite sequence of eigenstates of the
two-level Hamiltonian considered here, we characterize the corresponding
discrete geometric phase and compare its geometric dependence with that of
the continuous Berry phase arising from cyclic evolution.

In addition to the qubit formulation, a two-level Hilbert space can be represented in fermionic language by identifying its basis states with the vacuum and single-occupation sectors of a fermionic mode. Introducing a suitable Majorana encoding provides an equivalent operator representation in which
the Pauli algebra can be expressed through Majorana bilinears
\cite{Majorana1937, Kitaev2001,Alicea2012,Nayak2008}. Majorana operators and Majorana
bilinears are also central to the description of topological superconductors,
where spatially separated Majorana zero modes can emerge as low-energy
boundary excitations \cite{Kitaev2001,Alicea2012,LeijnseFlensberg2012}.
Their potential use for topological quantum information has motivated
extensive study in condensed-matter physics \cite{Nayak2008}.

This fermionic perspective also motivates extending the two-level construction to a parameter-dependent Bogoliubov--de Gennes family. In topological superconductors described by
Bogoliubov-de Gennes Hamiltonians, an effective two-level structure can
arise as a function of momentum over the Brillouin zone
\cite{Kitaev2001,Alicea2012, HasanKane2010}. For such momentum-dependent two-level bands, the geometric phase
associated with traversal of the Brillouin zone is the Berry (Zak) phase
\cite{Berry1984,Zak1989}, which, in symmetry-constrained
topological settings, can be related to global topological
characteristics such as winding numbers.

In this work, we start from a two-level mixing Hamiltonian inspired by
the formalism of neutral meson systems and map it to an effective qubit
representation. Within this framework, we analyze the geometric phase acquired during
cyclic parameter evolution and provide a geometric interpretation of the
\CP-like phase, including the reversal of the azimuthal orientation under
$\phi\rightarrow-\phi$ and its formal analogy with complex phase
conjugation in the mixing description. We further compute the
third-order Bargmann invariant and the associated discrete geometric phase,
highlighting its dependence on the \CP-like phase parameter. Here the designation ``\CP-like'' refers to the role of the complex phase
of the off-diagonal mixing amplitude within the Hermitian two-level
formulation and should not be interpreted as a rephasing-invariant measure
of physical \CP violation.  Building on these results, we show that the geometric contribution to the
cyclic evolution can be represented in the instantaneous eigenbasis as a
single-qubit $R_z$ phase operation.

We further express the two-level Hamiltonian in an equivalent fermionic
representation using Majorana operators, showing how the same two-level
operator structure and phase relations can be represented through
fermionic bilinears. This correspondence is representation-level and
does not imply that the neutral meson system realizes physical Majorana
modes. Finally, we extend the two-level structure to the Kitaev chain,
where the Bogoliubov--de Gennes Hamiltonian provides a momentum-dependent
family of two-level systems. For the real-parameter model considered
here, the relevant symmetry-constrained $d_y$--$d_z$ trajectory winds
around the origin, distinguishing the topological and trivial regimes.
For this model, the associated Berry (Zak) phase is quantized, while the
corresponding discrete Bargmann construction provides a characterization
of the Brillouin-zone geometric phase, approaching the Berry (Zak) phase
in the continuum limit.

The usefulness of this geometric perspective lies in providing a common
framework for related two-level structures across distinct physical
settings. These structures can be characterized through continuous
geometric phases, discrete BIs, geometric phase operations, and, in the
topological-superconducting realization, momentum-space winding. Taken
together, the constructions considered here provide a common perspective
on how phase structure in two-level Hamiltonians can be analyzed across
mixing and qubit dynamics, fermionic representations, and momentum-space
topology, while keeping their physical interpretations distinct.

\section{General Two-Level Mixing Formalism}
Neutral meson-antimeson systems provide a canonical physical realization of two-level quantum mixing. Let $|P^0\rangle$ and $|\overline{P^0}\rangle$ denote the flavor eigenstates spanning a two-dimensional Hilbert space, where $P^0-\overline{P^0}$ can represent $K^0-\overline{K^0}$, $B^0-\overline{B^0}$ or $D^0-\overline{D^0}$. Their time evolution can be described by an effective Hamiltonian acting on the state vector
\begin{align}
|\Psi(t)\rangle = a(t)\,|P^0\rangle + b(t)\,|\overline P^0\rangle,
\end{align}
through the Schr\"{o}dinger equation
\begin{align}
i\frac{d}{dt} \begin{pmatrix} a(t)\\ b(t) \end{pmatrix} = \mathcal{H} \begin{pmatrix} a(t)\\ b(t) \end{pmatrix}.
\end{align}
For the Hermitian, decay-free limit adopted in the present geometric
construction, the effective Hamiltonian takes the form \cite{Waldi2001,BigiSanda2009}
\begin{align}
\mathcal{H} = \begin{pmatrix} H_{11} & H_{12} \\ H_{12}^* & H_{22} \end{pmatrix}, \qquad H_{11},H_{22}\in\mathbb{R}, \quad H_{12}\in\mathbb{C}.
\end{align}
The off-diagonal element encodes flavor mixing and can be parameterized as
\begin{align}
H_{12} = |H_{12}|\,e^{-i\phi},
\end{align}
where $\phi$ parametrizes the complex phase structure of the mixing amplitude within the present two-level formulation and serves as the phase parameter whose geometric role is studied below. The Hamiltonian then becomes
\begin{align}
\mathcal{H} = \begin{pmatrix} H_{11} & |H_{12}|e^{-i\phi} \\ |H_{12}|e^{i\phi} & H_{22} \end{pmatrix}.
\end{align}
The eigenvalues are obtained from the characteristic equation $|\mathcal{H}-\lambda I|=0$, yielding
\begin{align}\label{eq:eigenvalues1}
\lambda_{\pm} = \frac{H_{11}+H_{22}}{2} \pm E,
\end{align}
where $E = \sqrt{\left(\frac{\delta}{2}\right)^2 + |H_{12}|^2}$ and $\delta = H_{11}-H_{22}$. 
The corresponding normalized eigenstates can be expressed in terms of mixing angle $\theta$ defined through
\begin{align}
\cos\theta = \frac{\delta/2}{E}, \qquad \sin\theta = \frac{|H_{12}|}{E},
\end{align}
with $0\leq\theta\leq\pi$. The corresponding eigenvectors can be written in this parametrization as 
\begin{subequations}
\begin{align}
|P_+\rangle &= \cos\frac{\theta}{2}\,|P^0\rangle + e^{i\phi}\sin\frac{\theta}{2}\,|\overline P^0\rangle,\\
|P_-\rangle &= -\sin\frac{\theta}{2}\,|P^0\rangle + e^{i\phi}\cos\frac{\theta}{2}\,|\overline P^0\rangle,
\end{align}
\end{subequations}
up to an overall phase convention.
Here, the phase $\phi$ parametrizes the complex phase of the off-diagonal mixing amplitude. Although the present Hamiltonian is restricted to a Hermitian, decay-free two-level form and therefore does not by itself encode the full rephasing-invariant structure of physical \CP violation, we retain this phase parameter to examine its geometric role within the chosen flavor-basis representation. The phase \(\phi\) is therefore treated as a \CP-like phase parameter whose geometric role is examined in the subsequent analysis, while \(\theta\) characterizes the mixing angle and hence the relative weight of the two flavor components \cite{Branco1999}.

The Hamiltonian can be conveniently written in the Pauli-matrix basis as \cite{Sakurai2021, NielsenChuang2010}
\begin{align}\label{eq:H_alpha}
\mathcal{H}=\alpha I+\vec{B}\cdot\vec{\sigma},
\end{align}
where $\alpha=\frac{H_{11}+H_{22}}{2}$, and
\begin{align}
\vec{B}=\big(|H_{12}|\cos\phi, |H_{12}|\sin\phi, {\delta \over 2}\big).
\end{align}
Here $\vec{B}$ denotes an effective parameter-space vector arising from the Hamiltonian decomposition and should not be interpreted as a physical magnetic field externally applied to a spin system.
 The spectrum is then given by \begin{align}\label{eq:eigenvalues2}
\lambda_{\pm}=\alpha\pm|\vec{B}|.
\end{align}

This representation makes explicit that the decay-free two-level mixing
problem can be characterized by an effective vector in a three-dimensional
parameter space. The parameters $|H_{12}|$, $\phi$, and $\delta$ determine the vector
$\vec B$, whose direction fixes the orientation of the corresponding
Hamiltonian eigenstates \cite{ShapereWilczek1989}. This geometric formulation will serve as the bridge to the qubit and geometric-phase analysis developed in the subsequent sections. 
\section{Mapping to an Effective Qubit Hamiltonian}
The geometric structure of the two-level mixing Hamiltonian admits a natural interpretation in terms of an effective qubit system. In particular, by comparing Eq.~(\ref{eq:eigenvalues1}) with Eq.~(\ref{eq:eigenvalues2}), one identifies
\begin{align}
|\vec{B}|=E.
\end{align}
Thus, the quantity $E$ introduced earlier as the half-splitting of the two eigenvalues is precisely the magnitude of $\vec{B}$ appearing in the Pauli decomposition. 

The flavor basis $\{|P^0\rangle, |\overline{P^0}\rangle\}$ spans a two-dimensional Hilbert space and may be identified with the computational basis $\{|0\rangle , |1\rangle\}$ of a qubit. Under this identification, the meson mixing Hamiltonian acts as an effective qubit Hamiltonian characterized by the Hamiltonian vector $\vec{B}$. The Hamiltonian eigenstates correspond to the eigenvectors of $\vec{B}\cdot \vec{\sigma}$, i.e. to the spinors aligned or anti-aligned with the direction of $\vec{B}$ on the Bloch sphere. Explicitly, writing $|\vec{B}|= \sqrt{\left(\frac{\delta}{2}\right)^2 + |H_{12}|^2}$, the Hamiltonian may be expressed as
\begin{align}
\mathcal{H}=\alpha I+|\vec{B}|\hat{n}\cdot \vec{\sigma},
\end{align}
where $\hat{n}={\vec{B}\over |\vec{B}|}$.
 The eigenstates are therefore the Bloch spinors pointing along $\pm \hat{n}$, and the level splitting equals $2|\vec{B}|=2E$. The previously introduced mixing angle $\theta$ is identified with the polar angle of $\hat{n}$, while $\phi$ defines its azimuthal angle in the $xy$-plane of parameter space. 

Thus, the mixing parameters $|H_{12}|$, $\phi$, and $\delta$ specify a
point in the coefficient space of the effective vector $\vec{B}$, whose
normalized direction $\hat{n}$ determines the corresponding point on the
Bloch sphere. Variations of the parameters that change the direction of
$\vec{B}$ therefore induce changes in the orientation of $\hat{n}$, while
variations that only rescale $|\vec{B}|$ leave this orientation unchanged.
The resulting unitary evolution generated by the Hermitian mixing
Hamiltonian is formally analogous to that of a spin-$1/2$ system in an
effective magnetic field, while $\vec{B}$ here remains a Hamiltonian
parameter-space vector rather than a physical applied field.

This mapping requires no additional dynamical assumptions beyond the Hermitian two-level approximation introduced earlier, and follows directly from the two-dimensional structure of the flavor Hilbert space. This exact two-level algebraic correspondence will serve as the foundation for the geometric analysis developed in the subsequent
sections, where parameter-space evolution of $\vec{B}$ will be examined.
\section{Geometric Phase in the Qubit Representation of the Mixing Hamiltonian}
In this section, we analyze the geometric phase within the qubit representation of the mixing Hamiltonian. Under the mapping established in the previous section between the neutral meson mixing Hamiltonian and an effective two-level qubit Hamiltonian, the flavor states are identified with qubit basis states according to $|P^0\rangle \leftrightarrow |0\rangle$, $|\overline{P^0}\rangle \leftrightarrow |1\rangle$. In the computational basis  $\{|0\rangle, |1\rangle\}$ the normalized eigenstates are
\begin{subequations}
\begin{align}
  |\psi_{+}\rangle &= \cos\frac{\theta}{2}\,|0\rangle + e^{i\phi}\sin\frac{\theta}{2}\,|1\rangle,\label{psi+}\\
  |\psi_{-}\rangle &= -\sin\frac{\theta}{2}\,|0\rangle + e^{i\phi}\cos\frac{\theta}{2}\,|1\rangle\label{psi-}.
\end{align}
\end{subequations}
The phase convention chosen for $|\psi_-\rangle$ provides a convenient
smooth representative for the subsequent geometric analysis. As usual,
the geometric phase is defined modulo $2\pi$, so equivalent representatives
may arise from different single-valued gauge choices for the eigenstates. These states correspond to Bloch spinors aligned and anti-aligned with the unit vector $\hat{n} = \vec{B}/|\vec{B}|$ \cite{Sakurai2021}. 
For a normalized eigenstate undergoing adiabatic cyclic evolution in parameter space, the geometric phase is defined by \cite{Berry1984}
\begin{align}
\gamma_\pm = i\oint \langle \psi_\pm | d\psi_\pm \rangle.
\end{align}
Considering the state $|\psi_+\rangle$ we have
\begin{align}
d|\psi_+\rangle = -\frac{1}{2}\sin\frac{\theta}{2} d\theta\,|0\rangle + e^{i\phi} \left( \frac{1}{2}\cos\frac{\theta}{2} d\theta + i\sin\frac{\theta}{2} d\phi \right)|1\rangle.
\end{align}
Substituting and simplifying, one obtains
\begin{align}
\langle \psi_+ | d\psi_+ \rangle = i\,\frac{1-\cos\theta}{2}\, d\phi.
\end{align}
For a cyclic evolution generated by varying the phase parameter $\phi$ from $0$ to $2\pi$, while keeping $|H_{12}|$ and $\delta$ (and hence $\theta$) fixed, the geometric phase becomes
\begin{align}\label{GP1}
\gamma_{g+} = -\pi(1-\cos\theta).
\end{align}
This equals minus one half of the solid angle subtended by the closed trajectory of $\hat{n}$ on the Bloch sphere \cite{Berry1984, AharonovAnandan1987}.
Repeating the same calculation for the orthogonal eigenstate $|\psi_-\rangle$ gives the geometric phase
 \begin{align}\label{GP2}
\gamma_{g-}
&=-\pi(1+\cos\theta)
\nonumber\\
&\equiv +\pi(1-\cos\theta)
\qquad (\mathrm{mod}\;2\pi).
\end{align}
The two eigenstates acquire geometric phases that are equal in magnitude
and opposite in sign,
$\gamma_{g-}\equiv-\gamma_{g+}\;(\mathrm{mod}\;2\pi)$.
This reflects the opposite orientations of the corresponding eigenstates
on the Bloch sphere.
In terms of the diagonal splitting $\delta$ and the off-diagonal mixing amplitude $H_{12}$, the geometric phase can be explicitly written as 
\begin{align}\label{GPpm}
\gamma_{g\pm}
&\equiv
\mp\pi
\left(
1-\frac{\delta}
{\sqrt{\delta^2+4|H_{12}|^2}}
\right)
\qquad (\mathrm{mod}\;2\pi).
\end{align}
This expression shows that the geometric phase is determined by the
relative magnitude and sign of the diagonal splitting $\delta$ and the
off-diagonal mixing amplitude $|H_{12}|$. The phase therefore encodes geometric information about the structure of the mixing Hamiltonian and, within the adiabatic regime, is independent of the rate of traversal of the parameter-space path.
 
The geometric phases derived in Eqs.(\ref{GP1}) and (\ref{GP2}) are invariant modulo $2\pi$ under single-valued phase redefinitions of the eigenstates, while the fixed-$\theta$ cyclic phase is unchanged under the corresponding constant phase redefinition of the computational basis. Consider first a phase transformation of the instantaneous eigenstates,
$|\psi_\pm\rangle \rightarrow e^{i\chi} |\psi_\pm\rangle$, where $\chi$ is a smooth function along the closed path. The Berry connection transforms as
$\langle \psi_\pm | d\psi_\pm \rangle \rightarrow \langle \psi_\pm | d\psi_\pm \rangle + i\,d\chi$. For a single-valued gauge transformation, $\oint d\chi=2\pi n$ with $n\in\mathbb{Z}$, so the geometric phase changes only by an integer multiple of $2\pi$. It is therefore gauge-invariant modulo $2\pi$. Next, consider a constant redefinition of the computational basis,
$|0\rangle \rightarrow e^{i\alpha}|0\rangle$, $|1\rangle \rightarrow e^{i\beta}|1\rangle$, 
which shifts the phase convention associated with the matrix element
$H_{12}$ while leaving $|H_{12}|$, $\delta$, and hence $E$ and $\theta$
unchanged. For the class of cyclic evolutions considered here (fixed $\theta$ loops), the geometric phase depends only on $\theta$ and is therefore unaffected by this transformation. Thus, $\gamma_{g\pm}$ is invariant modulo $2\pi$ under gauge transformations of the state, while the fixed-$\theta$ cyclic geometric phase is unchanged under the corresponding phase redefinition of the computational basis.

To obtain a clearer picture of the phase, we discuss the zero- and
strong-mixing limits. In the zero-mixing limit
$|H_{12}|\rightarrow 0$, the mixing angle approaches
$\theta\rightarrow0$ for $\delta>0$ and
$\theta\rightarrow\pi$ for $\delta<0$. Accordingly, the geometric
phases become trivial modulo $2\pi$: for $\delta>0$,
$\gamma_{g+}\rightarrow0$ and $\gamma_{g-}\rightarrow0$, while for
$\delta<0$, $\gamma_{g+}\rightarrow-2\pi\equiv0$ and
$\gamma_{g-}\rightarrow0$. At exactly $|H_{12}|=0$, however, the phase
$\phi$ of the off-diagonal amplitude is undefined, so the cyclic
construction should be understood as a limiting procedure from nonzero
mixing. Thus, the geometric phase becomes trivial in the zero-mixing
limit, as expected for eigenstates approaching the flavor basis states. In the opposite regime, corresponding to strong mixing,
$|H_{12}|\gg|\delta|$, one obtains
$\theta\rightarrow\pi/2$. The eigenstates become equal-weight
superpositions of $|0\rangle$ and $|1\rangle$. In this limit,
$\gamma_{g+}\rightarrow-\pi$, while
$\gamma_{g-}\rightarrow+\pi\equiv-\pi\;(\mathrm{mod}\;2\pi)$.
Thus, within the fixed-$\theta$ cyclic family considered here, maximal
mixing yields a geometric phase of magnitude $\pi$ modulo $2\pi$. This geometric behavior motivates a deeper interpretation
of the phase structure, which we develop in the following section in terms
of a \CP-like geometric phase.

\section{Geometric Interpretation of the \CP-like Phase}
In this section, we examine the geometric meaning of the complex phase appearing in the off-diagonal mixing term of the Hamiltonian. In neutral meson mixing, complex phases in the mixing amplitudes play an important role in the description of \CP-violating phenomena \cite{Christenson1964, KobayashiMaskawa1973}. In the present Hermitian two-level formulation, we consider the analogous phase parameter appearing in the off-diagonal mixing term, $H_{12}=|H_{12}|e^{-i\phi}$, and refer to it as a \CP-like phase. Here, $\phi$ is interpreted in terms of the geometric role of the complex mixing structure in the chosen flavor basis, rather than as a parameter characterizing physical \CP violation in a rephasing-invariant manner.

In the Pauli decomposition $\mathcal{H}=\alpha I+\vec{B}\cdot\vec{\sigma}$, where $\alpha={{H_{11}+H_{22}}\over 2}$, and the effective vector in this framework is given by
$\vec{B}=\left(|H_{12}|\cos\phi,\, |H_{12}|\sin\phi,\,{\delta \over 2}\right)$.
The eigenstates correspond to spinors aligned and anti-aligned with the direction of $\vec{B}$. The mixing angle $\theta$ fixes the relative weight of the basis states $|0\rangle$ and $|1\rangle$, whereas the phase $\phi$ determines the orientation of the transverse component of $\vec{B}$ in the plane orthogonal to the diagonal splitting. 

\begin{figure}[H]
	\centering
	\includegraphics[width=0.45\linewidth]{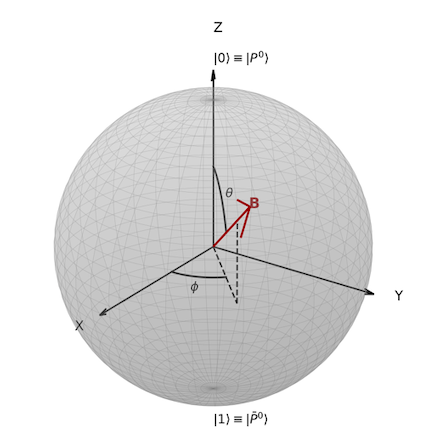}
	\caption{
		Geometric representation of the effective two–level Hamiltonian on the Bloch sphere.
		The basis states $|0\rangle \equiv |P^0\rangle$ and $|1\rangle \equiv |\overline P^0\rangle$ correspond to the north and south poles. The effective vector $\vec{B}$ specifies the direction $\hat n$ of the
        upper eigenstate, while the lower eigenstate is oriented along $-\hat n$. 
		The polar angle $\theta$ controls the mixing between the basis states, while the azimuthal angle $\phi$ specifies the orientation of the transverse component of $\vec{B}$ in the $xy$ plane. 
		The dashed projection onto the $xy$-plane illustrates the geometric interpretation of the \CP-like phase $\phi$ as the azimuthal orientation of the transverse Hamiltonian component.
	    }
	\label{fig:Bloch_sphere}
\end{figure}

Figure~\ref{fig:Bloch_sphere} provides a natural geometric visualization in terms of the Bloch sphere. In this representation, the poles of the sphere correspond to the basis states $|0\rangle$ and $|1\rangle$, while the eigenstates of the effective Hamiltonian lie along the direction specified by the unit vector $\hat{n}={\vec{B}\over |\vec{B}|}$. 
From this perspective, the \CP-like phase is not merely a phase parameter
in the matrix representation of the Hamiltonian, in the chosen flavor-basis
convention, it specifies the azimuthal orientation of the effective vector
$\vec{B}$ and hence the geometric orientation of the corresponding
eigenstates on the Bloch sphere.

When $\phi$ is varied adiabatically while the magnitudes $|H_{12}|$ and $\delta$ remain fixed, the direction $\hat{n}$ traces a closed curve at constant polar angle. This cyclic evolution reproduces the geometric phase derived in the previous section. The accumulated geometric phase is therefore controlled by the mixing angle
$\theta$, while variation of the \CP-like phase parameter $\phi$ governs the
azimuthal traversal that generates the closed trajectory in parameter space. In this way, the \CP-like phase $\phi$ acquires a clear geometric
interpretation as the parameter controlling the azimuthal traversal of the
closed path considered here, providing the geometric basis for the phase-gate
construction developed in the following sections.

\section{Orientation Reversal and Analogy with Meson Mixing Phases}
The geometric formulation developed above allows complex conjugation of the
off-diagonal mixing amplitude to be interpreted, within the present
two-level representation, as a transformation of the effective Bloch
vector $\vec{B}$. Since the orientation of the unit vector $\hat{n}$ determines the Hamiltonian eigenstates, reversal of the phase structure of the off-diagonal mixing amplitude corresponds to a reflection of this vector in parameter space. In particular, the phase parameter $\phi$, which appears in the off-diagonal
mixing element $H_{12}$, specifies, in the chosen flavor-basis convention,
the azimuthal orientation of the transverse component of $\vec{B}$ on the
Bloch sphere. For fixed $|H_{12}|$ and $\delta$, changing $\phi$ therefore modifies the azimuthal position of the Hamiltonian vector while leaving its polar angle unchanged. 

Consider the transformation $\phi\rightarrow-\phi$. Under this operation the components of $\vec{B}$ transform as
\begin{align}
\left(B_x,\,B_y,\,B_z\right)\rightarrow \left(B_x,-\,B_y,\,B_z\right),
\end{align}
which corresponds geometrically to a reflection of the Bloch vector across the $x-z$ plane, effectively reversing the orientation of its azimuthal coordinate.

\begin{figure}[H]
	\centering
	\includegraphics[width=0.45\linewidth]{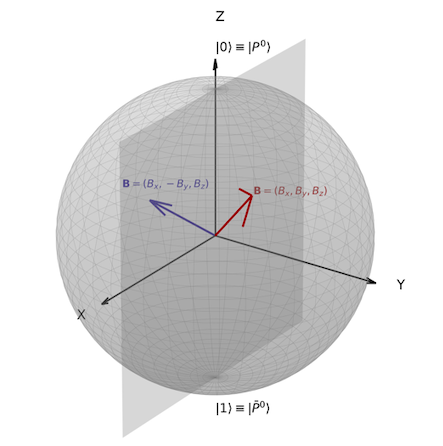}
	\caption{
		Reflection of the effective Hamiltonian vector under the transformation 
		$\phi\rightarrow -\phi$. The vectors $\vec B(\phi)$ and $\vec B(-\phi)$ are mirror images 
		across the $x-z$ plane while maintaining the same polar angle $\theta$.
	}
	\label{fig:Bloch_sphere2}
\end{figure}

Figure~\ref{fig:Bloch_sphere2} illustrates this reflection of the effective
Bloch vector under the transformation $\phi\rightarrow-\phi$. Since the geometric phase is proportional to the oriented solid angle
enclosed by the cyclic trajectory on the Bloch sphere, reversal of the
traversal direction changes the sign of the geometric phase. In the present
parameterization, the transformation $\phi\rightarrow-\phi$ maps the
cycle $\phi:0\rightarrow2\pi$ onto the oppositely oriented cycle
$\phi:0\rightarrow-2\pi$. Hence,
\begin{align}
    \gamma_{g\pm}\rightarrow-\gamma_{g\pm}
\qquad (\mathrm{mod}\;2\pi).
\end{align}

This behavior is analogous to the role of time-reversal symmetry in
neutral-meson mixing. Time reversal is represented by an antiunitary
operation, whose action involves complex conjugation in an appropriate
basis. In the present Hermitian two-level representation, complex
conjugation of the off-diagonal mixing element gives
$H_{12}\rightarrow H_{12}^{*}$. For
$H_{12}=|H_{12}|e^{-i\phi}$, this corresponds to
$\phi\rightarrow-\phi$, and hence to the reflection of the effective
Bloch vector described above. The resulting reversal of the oriented geometric phase therefore provides a geometric analogue of the phase-conjugation aspect associated with time reversal. It should not,
however, be interpreted as a direct implementation of, or a derivation
from, the physical time-reversal transformation of the neutral-meson
system.

\section{Bargmann Invariant and Discrete Geometric Phase}
A complementary probe of geometric structure in quantum mechanics is provided by the Bargmann invariants (BIs), which characterize geometric phase relations associated with closed sequences of quantum rays \cite{Bargmann1964, MukundaSimon1993}. BIs have previously been studied in the context of neutral meson systems, in particular, for the neutral kaon system, sequences involving flavor and mass eigenstates can be constructed to form BIs that encode geometric information about the mixing phases \cite{SangiriSarkar2023}. Motivated by this connection, we analyze the Bargmann structure within the present qubit representation of the two-level mixing Hamiltonian.

\begin{figure}[H]
 	\centering
 	\includegraphics[width=0.45\linewidth]{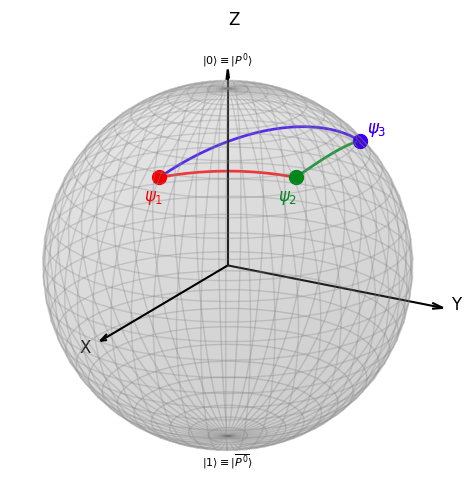}
 	\caption{
 		Bloch sphere representation of three states used to construct the third-order Bargmann invariant. The geodesic triangle connecting the states illustrates the discrete geometric phase associated with their cyclic sequence. The north and south poles correspond to the computational basis states $|0\rangle \equiv |P^0\rangle$ and $|1\rangle \equiv |\overline{P^0}\rangle$, respectively. The polar angle $\theta$ and azimuthal separation $\phi$ determine the positions of the states on the Bloch sphere, highlighting the geometric structure of the discrete phase.
 	}
 	\label{fig:BI_triangle}
 \end{figure}

For three normalized states forming a cyclic sequence of rays in the projective Hilbert space, the third-order BI is defined as 
\begin{align}
\Delta_3=\langle\psi_1|\psi_2\rangle\langle\psi_2|\psi_3\rangle\langle\psi_3|\psi_1\rangle,
\end{align}
 where no two consecutive states are orthogonal. Within the qubit formulation, the instantaneous eigenstates lie along the Bloch sphere direction specified by $(\theta,\phi)$. Consider three states obtained by only varying $\phi$: 
 \begin{align}
 |\psi_1\rangle = |\psi_+(\theta,0)\rangle, \quad |\psi_2\rangle = |\psi_+(\theta,\phi)\rangle, \quad |\psi_3\rangle = |\psi_+(\theta,2\phi)\rangle.
 \end{align}
 As illustrated in Fig.~\ref{fig:BI_triangle}, these states define, for nondegenerate choices of $\theta$ and $\phi$, a
closed sequence of rays that can be represented by a geodesic triangle on
the Bloch sphere, with its sides given by the shorter geodesic arcs
connecting the corresponding points.
 The overlaps between these states take the form
 \begin{align}
 \langle\psi(\theta,\phi_a)|\psi(\theta,\phi_b)\rangle= \cos^2\frac{\theta}{2} + e^{i(\phi_b-\phi_a)} \sin^2\frac{\theta}{2}.
 \end{align}
Evaluating the overlaps yields the BI
 \begin{align}
\Delta_3=\left(\cos^2{\theta\over 2}+e^{i\phi}\sin^2{\theta\over 2}\right)^2\left(\cos^2{\theta\over 2}+e^{-2i\phi}\sin^2{\theta\over 2}\right).
 \end{align}
 This expression makes explicit that the complex BI depends
on the mixing angle $\theta$ through the relative weights of the two
components, and its argument can therefore depend on the mixing geometry. The phase associated with the BI is defined as
  \begin{align}
   \gamma_{\Delta_3} = \arg(\Delta_3) = \mathrm{Im}\,\ln \Delta_3 \quad (\mathrm{mod}\ 2\pi),
  \end{align}
  which is invariant under rephasing of individual states and depends only on the closed sequence of rays.
  This construction admits a direct geometric interpretation: the three states define a closed path in projective Hilbert space corresponding to a geodesic triangle on the Bloch sphere, with the associated phase determined entirely by its geometry. In this sense, the BI provides a discrete analog of the continuous geometric phase, capturing the geometric structure without requiring a continuous dynamical evolution. 
  
  In the limiting cases $\theta\rightarrow0$ or $\theta\rightarrow\pi$,
the three rays collapse to the same basis-state ray and the discrete phase
becomes trivial. Likewise, for $\phi\rightarrow0$, the three states
coincide and the Bargmann phase vanishes. Thus, within the present parametrization, a nontrivial discrete geometric
contribution generally requires nontrivial mixing together with a nonzero
azimuthal separation.

The orientation-reversal property discussed in the preceding section also
has a direct counterpart at the level of the BI. Under the
azimuthal transformation $\phi\rightarrow-\phi$, one finds
\begin{align}
\Delta_3(\theta,-\phi)
&=
\left(
\cos^2\frac{\theta}{2}
+e^{-i\phi}\sin^2\frac{\theta}{2}
\right)^2
\left(
\cos^2\frac{\theta}{2}
+e^{2i\phi}\sin^2\frac{\theta}{2}
\right)
\nonumber\\
&=
\left[\Delta_3(\theta,\phi)\right]^*.
\end{align}
For nonzero $\Delta_3$, it is therefore natural to define the
orientation-reversal ratio
\begin{align}
\mathcal{R}_{\mathrm{rev}}(\theta,\phi)
\equiv
\frac{\Delta_3(\theta,\phi)}
{\Delta_3(\theta,-\phi)}.
\end{align}
Since the two invariants are complex conjugates, this ratio has unit
modulus and can be written as
\begin{align}
\left|\mathcal{R}_{\mathrm{rev}}\right|=1,
\qquad
\mathcal{R}_{\mathrm{rev}}
=
e^{2i\gamma_{\Delta_3}(\theta,\phi)}.
\end{align}
Thus, the ratio isolates the phase change associated with reversing the
azimuthal orientation of the discrete Bargmann construction, providing a
discrete counterpart to the orientation reversal of the continuous
geometric phase discussed in the preceding section.
For the particular three-state construction adopted here, the discrete
geometric phase depends explicitly on the azimuthal separation $\phi$
between the states. Since this separation is generated by varying the
azimuthal phase parameter $\phi$ of the mixing Hamiltonian within the
chosen parametrization, the corresponding Bargmann phase depends directly on this parameter through the azimuthal separations of the states. Thus, $\phi=0$ gives a trivial Bargmann
phase, whereas finite azimuthal separation generally produces a nonzero
phase whose value depends on both the separation and the mixing angle
$\theta$. Within the present two-level construction, this establishes a
direct geometric relation between the \CP-like phase parameter and the
corresponding discrete Bargmann phase, without identifying
$\gamma_{\Delta_3}$ with a rephasing-invariant measure of physical
\CP violation. This discrete formulation provides a natural bridge to
the holonomic phase construction, where cyclic evolution in parameter
space produces corresponding projective phase factors.

\section{Emergent Geometric Phase Gate}
The geometric structure derived in the previous sections motivates a formulation in terms of geometric phase and Abelian holonomy
\cite{ZanardiRasetti1999, Sjoqvist2012}. Under adiabatic variation of its parameters, the mixing Hamiltonian induces transport of its eigenstates in projective Hilbert space, corresponding to motion of the Bloch vector on the Bloch sphere.

The effective two-level Hamiltonian governing the system can be written as $\mathcal{H}=\alpha I+|\vec{B}|\,\hat{n}\cdot\vec{\sigma}$. The normalized eigenstates $|\psi_+\rangle$ and $|\psi_-\rangle$ correspond to spinors aligned with $\pm\hat{n}$. If the azimuthal angle $\phi$ varies adiabatically while $|H_{12}|$ and $\delta$ remain fixed, $\hat{n}$ traces a closed trajectory at constant polar angle $\theta$.

Under such cyclic evolution, each instantaneous eigenstate acquires
separate dynamical and geometric phase contributions. After completion of
the loop, the eigenstates return to themselves up to a phase factor
\begin{align}
|\psi_{\pm}(T)\rangle = e^{\,i(\gamma_{d\pm}+\gamma_{g\pm})} |\psi_{\pm}(0)\rangle,
\end{align} 
where $\gamma_{d\pm}$ are the corresponding dynamical phases and
$\gamma_{g\pm}$ are the geometric phases. For the class of cyclic evolutions considered here, we have $\gamma_{g+}=-\pi(1-\cos\theta)$ and  $\gamma_{g-}=+\pi(1-\cos\theta)$. Consequently, in the ordered eigenbasis
$\{|\psi_+\rangle,|\psi_-\rangle\}$, the geometric contribution to the
evolution operator becomes
\begin{align}
U_g=
\begin{pmatrix}
e^{-i\pi(1-\cos\theta)} & 0\\
0 & e^{\,i\pi(1-\cos\theta)}
\end{pmatrix},
\end{align}
which can be written as
\begin{align}
U_g=\exp\!\left[-\,i\,\pi(1-\cos\theta)\,\sigma_z\right].
\end{align}
Equivalently, for each instantaneous eigenstate, the geometric phase can be expressed through the Berry connection. For an adiabatic cyclic evolution along a closed path $\mathcal{C}$ in parameter space,
\begin{align}
\gamma_{g\pm}(\mathcal{C})
=
\oint_\mathcal{C}\mathcal{A}_{\pm}(\lambda)\cdot d\lambda,
\end{align}
where
\begin{align}
\mathcal{A}_{\pm}(\lambda)
=
i\langle\psi_{\pm}(\lambda)|
\nabla_{\lambda}\psi_{\pm}(\lambda)\rangle
\end{align}
is the Berry connection for the corresponding eigenstate. The geometric evolution operator in the two-dimensional eigenbasis is then constructed from the geometric phases of the two eigenstates. For the present trajectory the Bloch vector sweeps a closed loop that subtends the solid angle $\Omega=2\pi(1-\cos\theta)$. The geometric phase then takes the form
\begin{align}
\gamma_{g\pm}=\mp\,\frac{\Omega}{2},
\end{align}
so that the geometric evolution operator becomes
\begin{align}\label{eq:Ug}
U_g=\exp\!\left[-\,i\,\frac{\Omega}{2}\,\sigma_z\right].
\end{align}
Using Eq.~(\ref{GPpm}) we can express the corresponding geometric
holonomy as
\begin{align}
U_g = \exp \left[- i\pi \left( 1- \frac{\delta}{\sqrt{\delta^2+4|H_{12}|^2}} \right) \sigma_z \right].
\end{align}
This unitary represents a single--qubit geometric phase gate whose
rotation angle is determined by the relative magnitude and sign of the
diagonal splitting $\delta$ and the off-diagonal mixing amplitude
$|H_{12}|$.

In the instantaneous eigenbasis $\{|\psi_+\rangle,|\psi_-\rangle\}$, the action of the gate corresponds to a relative phase rotation,
\begin{align}
|\psi_+\rangle \rightarrow e^{-i\Omega/2}\,|\psi_+\rangle,
\qquad
|\psi_-\rangle \rightarrow e^{\,i\Omega/2}\,|\psi_-\rangle .
\end{align}
Thus, cyclic geometric transport of the effective Bloch vector realizes a geometric phase gate whose magnitude depends on the geometry of the trajectory in parameter space. The meson-qubit correspondence therefore provides a formal mapping from
the geometric structure of flavor mixing to a single-qubit geometric phase
gate, whose rotation angle, for the cyclic path considered here, is determined
by the parameters of the mixing Hamiltonian.

From the quantum-computational perspective, the geometric phase gate
derived in Eq.~(\ref{eq:Ug}) is precisely a single-qubit $R_z$ rotation,
\begin{align}
U_g=R_z(\Omega),
\end{align}
where $R_z(\Omega)=\exp(-i\Omega\sigma_z/2)$.
This establishes a formal mapping between the mixing parameters and the
corresponding single-qubit geometric rotation angle. If independent
$R_x$ and $R_y$ rotations are available, the resulting set of single-qubit
rotations is sufficient to generate arbitrary single-qubit unitaries
(up to an overall phase) \cite{Barenco1995}. Thus, the present construction
provides a geometric realization of the $R_z$ component rather than, by
itself, a complete universal gate set. This suggests a conceptual
connection between particle-mixing dynamics and geometric gate design for
quantum computation. The corresponding operator-level formulation is
developed in the following section in terms of Majorana fermions.

\section{Majorana Representation of the Two-Level Mixing Hamiltonian and its Geometric Interpretation}
The two-level structure of the neutral meson mixing Hamiltonian admits a formally equivalent formulation in terms of fermionic degrees of freedom, which provides an alternative perspective on the geometric structure developed in the preceding sections. Majorana operators are Hermitian, and suitable bilinears of a sufficient set of Majorana modes can represent the Pauli operators appearing in the two-level Hamiltonian \cite{Kitaev2001,Alicea2012,Nayak2008}. Within this representation, the two-level operators underlying the
geometric-phase construction can be expressed through corresponding
Majorana bilinears, providing an operator-level representation of the
geometric structure. This provides a representation-level correspondence without requiring additional dynamical assumptions about the neutral meson system. Such Majorana representations play a central role in condensed matter
physics, particularly in topological superconductors where Majorana modes
can emerge as effective low-energy excitations and are of interest for
topological quantum information processing
\cite{Stanescu2011, LeijnseFlensberg2012(2), ElliottFranz2015, SatoFujimoto2016}.

We introduce fermionic operators $c$ and $c^\dagger$ satisfying
$\{c, c^\dagger\} = 1$,
and define the Majorana operators
\begin{align}\label{eq:gamma12}
\gamma_1 = c + c^\dagger, \qquad \gamma_2 = -i(c - c^\dagger),
\end{align}
which obey
\begin{align}
\{\gamma_i, \gamma_j\} = 2\delta_{ij}, \qquad \gamma_i^\dagger = \gamma_i.
\end{align}
These operators generate the Clifford algebra associated with a single
fermionic mode, whose two-dimensional Fock space is isomorphic, at the
level of Hilbert-space structure, to the two-level Hilbert space spanned
earlier by the flavor basis $\{|P^0\rangle,|\overline{P^0}\rangle\}$, or
equivalently the qubit basis $\{|0\rangle,|1\rangle\}$.

The single-mode Majorana representation provides an explicit realization
of one Pauli operator. Indeed, starting from the definitions of
Eq.~(\ref{eq:gamma12}), one readily verifies that
\begin{align}
i\gamma_1\gamma_2=(c+c^\dagger)(c-c^\dagger)\nonumber.
\end{align}
Expanding the product and using $\{c, c^\dagger\} = 1$, one obtains $(c+c^\dagger)(c-c^\dagger)=2c^\dagger c-1$. 
Thus, $i\gamma_1\gamma_2=2c^\dagger c-1$,
and, with the convention
$\sigma_z=|0\rangle\langle0|-|1\rangle\langle1|$,
\begin{align}
\sigma_z=-i\gamma_1\gamma_2.
\end{align}
The remaining Pauli operators require a larger Majorana encoding. A
convenient realization uses four Majorana operators
$\Gamma_1,\Gamma_2,\Gamma_3,\Gamma_4$. These $\Gamma_i$ denote a distinct four-Majorana encoding introduced for
the purpose of representing the full Pauli algebra and they should not be
identified with the two Majorana operators $\gamma_1,\gamma_2$ introduced
above. The corresponding four-dimensional
fermionic Fock space decomposes into two-dimensional even- and odd-parity
sectors, restricting to either fixed-parity sector gives an encoded
two-level Hilbert space. Within such a sector, suitable Majorana bilinears can be identified with the three Pauli operators. One convenient choice, up to signs determined by the selected parity sector and encoding convention, is
\begin{align}
\sigma_x &= i\Gamma_1\Gamma_2,\qquad
\sigma_y = i\Gamma_2\Gamma_3,\qquad
\sigma_z = i\Gamma_1\Gamma_3.
\end{align}
With a consistent choice of these signs within a fixed-parity sector, the corresponding bilinears reproduce the Pauli algebra in the encoded two-level subspace. These bilinears preserve total fermion parity and, upon restriction to a
fixed-parity sector, reproduce the Pauli algebra within the corresponding
encoded two-level subspace. Accordingly, the
traceless part $\vec{B}\cdot\vec{\sigma}$ of the two-level Hamiltonian can
be represented by quadratic Majorana bilinears. For the sign convention adopted above,
\begin{align}
\vec{B}\cdot\vec{\sigma}
=
iB_x\Gamma_1\Gamma_2
+iB_y\Gamma_2\Gamma_3
+iB_z\Gamma_1\Gamma_3.
\end{align}
More generally, the Hamiltonian can be written in quadratic Majorana form as
\begin{align}
\mathcal{H}
=
\alpha I+\frac{i}{2}\sum_{i,j}A_{ij}\Gamma_i\Gamma_j,
\end{align}
where $A_{ij}$ is a real antisymmetric matrix. Its independent entries are
fixed by the components of $\vec B$ once the Majorana encoding and parity
sector are specified. Thus, within the chosen fixed-parity sector, the
mixing parameters of the original two-level Hamiltonian admit an
equivalent representation in terms of Majorana-bilinear couplings.

This correspondence also extends the geometric interpretation to the
operator level. Variations of the parameters $|H_{12}|$, $\phi$, and
$\delta$ modify the coefficients of the corresponding Majorana bilinears.
In particular, the azimuthal variation of $\phi$ changes the relative
contributions of the bilinears representing $\sigma_x$ and $\sigma_y$,
providing an operator-level representation of the parameter-space
evolution without introducing a new dynamical mechanism for the phase. Within the four-Majorana encoding, the geometric phase gate obtained in
the preceding section can itself be written in Majorana-bilinear form.
Since $\sigma_z=i\Gamma_1\Gamma_3$, one has
\begin{align}
U_g
&=
\exp\!\left(-\frac{i\Omega}{2}\sigma_z\right)
\nonumber\\
&=
\exp\!\left(\frac{\Omega}{2}\Gamma_1\Gamma_3\right).
\end{align}
Using the expression for $\Omega$ obtained from the mixing Hamiltonian,
this can equivalently be written as
\begin{align}
U_g
=
\exp\!\left[
\pi\left(
1-\frac{\delta}
{\sqrt{\delta^2+4|H_{12}|^2}}
\right)
\Gamma_1\Gamma_3
\right].
\end{align}
Thus, the geometric phase gate admits an operator-level representation
generated by a Majorana bilinear within the chosen encoded two-level
subspace.

In physical topological-superconductor systems supporting spatially
separated Majorana zero modes, Majorana-bilinear unitaries also arise in
the description of Majorana exchange \cite{Ivanov2001}. For a standard
exchange convention, the exchange of two Majorana modes is represented by
\begin{align}
U_{12}=\exp\!\left(\frac{\pi}{4}\Gamma_1\Gamma_2\right).
\end{align}
This operation is distinct from the geometric phase gate derived above:
$U_g$ is the geometric transformation obtained within the present
two-level encoding, whereas $U_{12}$ describes the exchange of physical
Majorana modes in a topological-superconductor setting. The neutral meson
system does not realize such physical braiding. The relevance here is
therefore algebraic, namely that both transformations are generated by
Majorana bilinears, providing a connection between the two-level
geometric construction and the physical Majorana framework considered in
the following section.

The fermionic representation also gives a direct interpretation of the
phase structure of the two-level eigenstates. With
\begin{align}
|1\rangle=c^\dagger|0\rangle,
\end{align}
the state
$|\psi_+(\theta,\phi)\rangle$ corresponds to a coherent superposition of the
vacuum and single-particle sectors, with $\phi$ appearing as their relative
phase. The geometric and Bargmann phases discussed above can, within this
representation, be related to the relative phase structure between the two
occupation sectors. Within the present two-level
correspondence, this provides a fermionic representation-level analogue of
the \CP-like phase introduced earlier: $\phi$ controls the relative phase
between the two basis sectors, rather than representing a physical
\CP transformation of the fermionic system.

The Majorana formulation thus provides an alternative fermionic
representation of the geometric structure of the two-level mixing
Hamiltonian. The qubit description expresses this structure through the
orientation of the effective Bloch vector, while the Majorana description
encodes the corresponding two-level operators through fermionic bilinears
and the associated relative phase structure. This establishes the
representation-level connection between particle mixing, qubit dynamics,
and Majorana operators, while leaving the physical realization of Majorana
modes and their topological properties to the condensed-matter setting
developed in the following section.

\section{Extension to Topological Superconducting Systems and Majorana Modes}
The Majorana representation developed in the previous section shows that the
two-level mixing Hamiltonian admits a formulation in terms of real fermionic
operators, providing an operator-level perspective on the geometric structure
developed above. This provides a natural extension from a single two-level Hamiltonian to a continuous family of such Hamiltonians. In this section, we
realize this extension through the Kitaev chain and show how the geometric
phase structure of the two-level description acquires a topological
character through the momentum-space winding of its BdG
Hamiltonian.
\subsection{Kitaev Chain as a Momentum-Space Realization of Two-Level Mixing}
We recall the effective Hamiltonian of the two-level system introduced in Eq.~(\ref{eq:H_alpha}):
\begin{align}
\mathcal{H}=\alpha I+\vec{B}\cdot\vec{\sigma}.\nonumber
\end{align}
This Hamiltonian can be promoted to a parameter-dependent family in the following way
\begin{align}
\mathcal{H}(\lambda)=\alpha(\lambda) I+\vec{B}(\lambda)\cdot\vec{\sigma},
\end{align}
where $\lambda$ denotes a continuous parameter labeling the Hamiltonian.
When $\lambda$ is varied cyclically, the corresponding Hamiltonian traces a
closed trajectory in parameter space. For a periodic parameter, such as the crystal momentum $k$ of a lattice
system, this trajectory is naturally closed.

A concrete realization of this structure is provided by the one-dimensional topological superconductor introduced by Kitaev \cite{Kitaev2001}, which describes a chain of spinless fermions with nearest-neighbor hopping and p-wave superconducting pairing.

The real-space Hamiltonian of the model is given by
\begin{align}
H = -\mu \sum_j c_j^\dagger c_j
-\; t \sum_j \big(c_j^\dagger c_{j+1} + \mathrm{h.c.}\big)
+\frac{\Delta}{2}\sum_j
\big(c_j c_{j+1} + \mathrm{h.c.}\big),
\end{align}
where $c_j^\dagger$ and $c_j$ are the fermionic creation and annihilation
operators at lattice site $j$. The parameter $\mu$ denotes the chemical
potential, $t>0$ is the nearest-neighbour hopping amplitude, and $\Delta$
is the superconducting pairing parameter. The factor $1/2$ in the pairing
term is a convention adopted here to fix the normalization of $\Delta$.
The overall sign of the pairing term is fixed by the Fourier and Nambu
conventions adopted below.

\begin{figure}[H]
	\centering
	\includegraphics[width=0.55\linewidth]{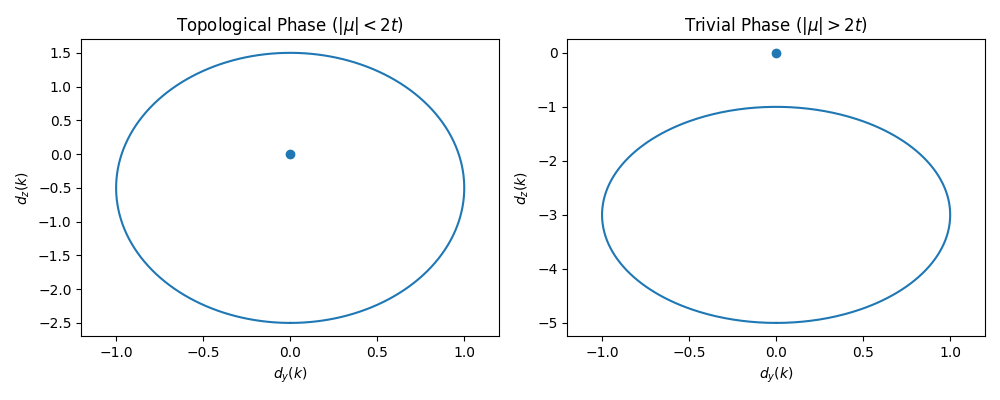}
	\caption{
Momentum-space representation of the Bogoliubov--de Gennes Hamiltonian
$H(k)=\vec d(k)\cdot\vec\sigma$ for the Kitaev chain in the
$d_y$--$d_z$ plane. The trajectory
$\vec d(k)=(0,\Delta\sin k,-\mu-2t\cos k)$ is shown for (left) the
topological regime $(|\mu|<2t)$ and (right) the trivial regime
$(|\mu|>2t)$. The origin $(0,0)$ marks the degeneracy point in the
$d_y$--$d_z$ plane, corresponding to $|\vec d(k)|=0$. The bulk gap closes when the momentum-space trajectory
passes through this point. For $t>0$ and $\Delta>0$, the topological
regime has a nonzero winding number, taken here as $\nu=1$ according to
the chosen orientation convention, whereas the trivial regime has
$\nu=0$. The figure illustrates the geometric characterization of the
topological distinction through the winding of $\vec d(k)$ around the
origin over the Brillouin zone.
	}
	\label{fig:Topological_and_total_phase}
\end{figure}

To expose its two-level structure, it is convenient to work in momentum space. Introducing the Fourier transform
\begin{align}
c_j = \frac{1}{\sqrt{N}} \sum_k e^{ikj} c_k,
\end{align}
and defining the Nambu spinor 
\begin{align}
\Psi_k = \begin{pmatrix} c_k \\ c_{-k}^\dagger \end{pmatrix},
\end{align}
the Hamiltonian can be written in BdG form as
\begin{align}
H = \frac{1}{2}\sum_k \Psi_k^\dagger H(k)\,\Psi_k + \mathrm{const.}
\end{align}
where $H(k)$ is a $2\times2$ Hermitian matrix acting in particle–hole space \cite{HasanKane2010}.
For the Kitaev chain, this matrix takes the form
\begin{align}
H(k) =
\begin{pmatrix}
-\mu-2t\cos k & -i\,\Delta\sin k\\
i\,\Delta\sin k & \mu+2t\cos k
\end{pmatrix}.
\end{align}
At this stage, one observes that $H(k)$ is a two-level Hamiltonian for each momentum k. Using the general decomposition of $2\times2$ Hermitian matrices in the Pauli basis discussed earlier, it can be expressed as
\begin{align}
H(k)=\vec{d}(k)\cdot\vec{\sigma},
\end{align}
with
\begin{align}
\vec{d}(k)=\big(0,\ \Delta \sin k,\ -\mu - 2t\cos k\big).
\end{align}
This geometric structure is illustrated in Fig.~(\ref{fig:Topological_and_total_phase}).

This representation makes explicit that the Kitaev chain is described by a
continuous family of effective two-level BdG Hamiltonians, each labeled by
the momentum $k$. The Brillouin zone $k\in[-\pi,\pi]$ therefore plays the role of the parameter space introduced previously.
The correspondence with the general parameter-dependent Hamiltonian
\begin{align}
\mathcal{H}(\lambda)=\alpha(\lambda)I+\vec{B}(\lambda)\cdot\vec{\sigma}
\end{align}
now becomes transparent. The momentum variable $k$ serves as the parameter $\lambda$, while the vector $\vec{d}(k)$ plays the role of the effective parameter-space vector
$\vec{B}(\lambda)$. In contrast to the earlier sections, where the parameter variation was introduced through the phase $\phi$, here the parameter arises intrinsically from the band structure of the system.
This identification provides a direct bridge between the two frameworks: the geometric structure associated with the vector $\vec B$ in the qubit
representation is realized in the momentum-dependent coefficient vector
$\vec d(k)$ of the BdG Hamiltonian. As $k$ varies across the Brillouin zone, $\vec{d}(k)$ traces a closed trajectory in parameter space, thereby enabling a geometric and topological interpretation of the Hamiltonian analogous to that developed in the context of two-level mixing.

The Kitaev chain represents a minimal realization of this topological mechanism, while related Majorana platforms arise in more realistic superconducting systems with additional microscopic structure. For example, topological superconductivity and Majorana bound states have been studied at the $LaAlO_3/SrTiO_3$ interface, where spin-orbit coupling and external Zeeman fields have been investigated as ingredients for realizing topological superconducting regimes supporting Majorana bound states \cite{MohantaTaraphder2014}. Such systems provide examples of more complex Bogoliubov--de Gennes
platforms in which topological superconductivity and Majorana bound states
can arise beyond the minimal Kitaev-chain description.
\subsection{Geometric Phases, Bargmann Invariants, and Topological Phase Quantization}

The momentum-dependent two-level structure of the Kitaev chain provides a
concrete extension of the geometric framework developed above. For each
momentum $k$, the Bogoliubov--de Gennes Hamiltonian defines an effective
two-level system with coefficient vector
$\vec d(k)=(0,\Delta\sin k,-\mu-2t\cos k)$. As $k$ traverses the Brillouin
zone, this vector traces a closed trajectory in the $d_y$--$d_z$ plane,
whose winding around the origin distinguishes the topological and trivial
regimes. The transition between these regimes occurs when the bulk gap closes, $|\vec d(k)|=0$, which, for $\Delta\neq0$, occurs at $|\mu|=2t$. In the topological regime, the corresponding open Kitaev chain supports
Majorana boundary modes 
\cite{Kitaev2001,Alicea2012}.

The geometric phase associated with the momentum-space parameter cycle can
be described directly in terms of a discrete Bargmann construction. In analogy
with the finite-state BI introduced earlier, consider a
partition of the Brillouin zone into points
$k_1,\ldots,k_N$, with $k_{N+1}=k_1$, and define
\begin{align}
\Gamma_N =
\arg\left[
\prod_{i=1}^{N}
\langle u(k_i)|u(k_{i+1})\rangle
\right],
\end{align}
where $|u(k_i)\rangle$ denotes the chosen BdG eigenstate and consecutive
states are assumed to be nonorthogonal. The cyclic product is invariant
under independent phase redefinitions of the individual states and
therefore depends only on the sequence of rays traced across the
Brillouin zone.

In the continuum limit in which the maximum separation between consecutive
points tends to zero, the discrete geometric phase approaches the Berry
(Zak) phase of the occupied BdG band,
\begin{align}
\lim_{\max_i|k_{i+1}-k_i|\to0}\Gamma_N
=
\gamma
\qquad (\mathrm{mod}\;2\pi).
\end{align}
Thus, the Bargmann construction provides a discrete counterpart of the
continuous geometric phase, extending the finite-state geometric
description developed earlier to a momentum-resolved family of two-level
states.

For the real-parameter Kitaev chain considered here, with $\Delta\neq0$,
the symmetry-constrained planar structure of $\vec d(k)$ leads to a
quantized Berry (Zak) phase in the gapped regimes. The corresponding
winding number $\nu$ counts the number of times the trajectory
$(d_y(k),d_z(k))$ winds around the origin as $k$ traverses the Brillouin
zone. With the winding convention adopted for
Fig.~\ref{fig:Topological_and_total_phase}, the Berry (Zak) phase is
\begin{align}
\gamma=\pi\nu
\qquad (\mathrm{mod}\;2\pi),
\end{align}
where, for the present Kitaev chain, $\nu=1$ and $\nu=0$ distinguish the topological and trivial regimes, respectively, under the convention adopted here. Consequently, as the momentum
partition is refined, the finite Bargmann phase $\Gamma_N$ approaches
the quantized geometric phase associated with the underlying topological
sector.
This establishes a direct connection between the finite-state Bargmann
geometry developed for the two-level mixing Hamiltonian and the global
momentum-space geometry of the Kitaev chain. In the former case, the
BI characterizes geometric phase relations among a finite
sequence of quantum states; in the latter, the same construction extends
over the Brillouin zone, where the global phase structure is tied to the
winding of the BdG Hamiltonian. The resulting correspondence provides a
discrete geometric bridge between two-level mixing, geometric phase, and
the topological characterization of the Kitaev chain.

\section{Summary}
In this work, we investigated the geometric structure of a Hermitian
two-level mixing Hamiltonian inspired by neutral meson mixing and mapped it
to an effective qubit description. The Pauli representation associates the
Hamiltonian with an effective Bloch vector, whose cyclic adiabatic evolution
generates a Berry phase determined by the solid angle enclosed on the Bloch
sphere. Within the fixed-$\theta$ family considered here, this phase is
controlled by the diagonal splitting and the magnitude of the off-diagonal
mixing amplitude.

The complex phase of the off-diagonal term was interpreted geometrically as
the azimuthal parameter of the effective Bloch vector. Its reversal changes
the orientation of the corresponding path and reverses the geometric phase.
We refer to this quantity as a \CP-like phase parameter, emphasizing that,
within the present Hermitian formulation, it is not identified with a
rephasing-invariant measure of physical \CP violation.

A complementary characterization of the geometric structure is provided by
the third-order Bargmann invariant. Constructing a finite cyclic sequence of
eigenstates yields a rephasing-invariant complex quantity whose argument
defines a discrete geometric phase. Unlike the continuous Berry phase, this
construction characterizes the geometry of a closed sequence of rays through
their mutual overlaps. In the present construction, the Bargmann phase
depends on the mixing angle and the azimuthal separation between the states,
while orientation reversal is reflected through the corresponding
complex-conjugation relation between the Bargmann invariants.

The cyclic geometric evolution further gives rise to a single-qubit
geometric phase operation, represented in the instantaneous eigenbasis as
an $R_z$ rotation. The corresponding two-level operator structure also
admits a fermionic representation in terms of Majorana bilinears, providing
a representation-level connection between the mixing Hamiltonian, qubit
dynamics, and a Majorana encoding without implying physical Majorana modes
in the neutral meson system.

Finally, we extended the geometric analysis to the momentum-dependent
two-level Bogoliubov--de Gennes Hamiltonian of the Kitaev chain. In this
minimal topological setting, the winding of the symmetry-constrained
coefficient vector distinguishes the topological and trivial regimes. The
Bargmann construction provides a complementary discrete description of the
Brillouin-zone geometric phase, approaching the corresponding Berry (Zak)
phase in the continuum limit.

Overall, the results provide a common geometric framework for examining
continuous and discrete phases, geometric qubit operations, fermionic
representations, and momentum-space topology within related two-level
Hamiltonian structures, while keeping their distinct physical
interpretations separate. The framework may be extended to more general
Bogoliubov--de Gennes models, non-Hermitian or open two-level systems, and
more general parameter-space trajectories.

%{\it Acknowledgement :} S. S. would like to thank MoE, Government of India for the research fellowship.


\begin{thebibliography}{99}
	\bibitem{BigiSanda2009}
	I.~I.~Bigi and A.~I.~Sanda, 
	\textit{CP Violation}
	(Cambridge University Press, Cambridge, 2009).
	
	\bibitem{LeeOehmeYang1957}
	T.~D.~Lee, R.~Oehme, and C.~N.~Yang,
	Phys.\ Rev.\ \textbf{106}, 340 (1957).
	
	\bibitem{GellMannPais1955}
	M.~Gell-Mann, A. Pais,
	Phys. Rev. \textbf{97}, 1387 (1955).
	
	\bibitem{Christenson1964}
	J.~H.~Christenson \textit{et al.},
	Phys.\ Rev.\ Lett.\ \textbf{13}, 138 (1964).
	
	\bibitem{Wolfenstein1964}
	L.~Wolfenstein,
	Phys. Rev. Lett. \textbf{13}, 562 (1964).
	
	\bibitem{KobayashiMaskawa1973}
	M.~Kobayashi and T.~Maskawa,
	Prog.\ Theor.\ Phys.\ \textbf{49}, 652 (1973).
	
	\bibitem{Nir1999}
	Y.~Nir,
	\textit{Flavor physics and CP violation, 1998 European School Of High-Energy Physics, Proceedings}, \textbf{99}(4), 99 (1999).
	
	\bibitem{Sarkar2008}
	U.~Sarkar,
	\textit{Particle and Astroparticle Physics}
	(Taylor and Francis, 2008).
	
	\bibitem{Berry1984}
	M.~V.~Berry,
	Proc.\ R.\ Soc.\ Lond.\ A \textbf{392}, 45 (1984).
	
	\bibitem{Simon1983}
	B.~Simon, 
	Phys. Rev. Lett. \ \textbf{51}, 2167 (1983).
	
	\bibitem{Anandan1992}
	J.~Anandan,
	Nature, \textbf{360}, 307 (1992).
	
	\bibitem{ZanardiRasetti1999}
	P.~Zanardi, M.~Rasetti,
	Phys. Lett. A \ \textbf{264}, 94 (1999).
	
	\bibitem{Bargmann1964}
	V.~Bargmann,
	J.\ Math.\ Phys.\ \textbf{5}, 862 (1964).
	
	\bibitem{Rabei1999}
	E.~M.~Rabei, Arvind, N.~Mukunda, and R.~Simon,
	Phys.\ Rev.\ A \textbf{60}, 3397 (1999).

    \bibitem{MukundaSimon1993}
     N.~Mukunda and R.~Simon, 
     Ann. Phys. (N.Y.) \textbf{228}, 205 (1993).

    \bibitem{SangiriSarkar2023}
     S.~Sangiri and U.~Sarkar,
     Nucl.\ Phys.\ B \textbf{990}, 116169 (2023).

     \bibitem{Sangiri2026}
      S.~Sangiri,
      Eur. Phys. J. C \textbf{86}, 592 (2026). 

      \bibitem{Hetenyi2026}
      H.~Bal{\'a}zs,
      J. Phys.: Condens. Matter \textbf{38}, 153002 (2026).
	
	%\bibitem{Nielsen2010}
	%M.~A.~Nielsen, I.~L.~Chuang,
	%\textit{ Quantum Computation and Information},
	%(Cambridge University Press: NY, USA, 2010).
	
	\bibitem{Majorana1937}
	E.~Majorana,
	Nuovo Cim. \textbf{14}, 171 (1937).
	
	\bibitem{Kitaev2001}
	A.~Kitaev,
	Phys.-Usp. \textbf{44} 131(2001).
	
	\bibitem{Alicea2012}
	J.~Alicea,
	Rep. Prog. Phys. \textbf{77} 076501 (2012).
	
	\bibitem{Nayak2008}
	C.~Nayak, S.~H.~.Simon, A.~Stren,M.~Freedman and S.~D.~Sarma
	Rev. Mod. Phys. \textbf{80}, 1083 (2008).

    \bibitem{LeijnseFlensberg2012}
	M.~Leijnse and K.~Flensberg,
	Phys. Rev. B \textbf{86}, 134528 (2012).
	
	\bibitem{HasanKane2010}
	M.~Z.~Hasan and C.~L.~Kane,
	Rev. Mod. Phys. \textbf{82}, 3045 (2010).

    \bibitem{Zak1989}
    J.~Zak,
    Phys. Rev. Lett. \textbf{62}, 2747 (1989).
	
	\bibitem{Waldi2001}
	R.~Waldi,
	Prog.\ Part.\ Nucl.\ Phys.\ \textbf{47}, 1 (2001).
	
	\bibitem{Branco1999}
	G.~C.~Branco, L.~Lavoura, and J.~P.~Silva,
	\textit{CP Violation}
	(Oxford University Press, Oxford, 1999).
	
	\bibitem{Sakurai2021}
	J.~J.~Sakurai and B J.~Napolitano
	\textit{Modern Quantum Mechanics}
	(Cambridge University Press, 2010).
	
	\bibitem{NielsenChuang2010}
	M.~A.~Nielsen and I.~L.~Chuang,
	\textit{Quantum Computation and Quantum Information}
	(Cambridge University Press, 2010).
	
	\bibitem{ShapereWilczek1989}
	A.~Shapere and F.~Wilczek
	\textit{Geometric Phases in Physics}
	(World Scientific Publishing, 1989).
	
	\bibitem{AharonovAnandan1987}
	Y.~Aharonov and J.~Anandan,
	Phys.\ Rev.\ Lett.\ \textbf{58}, 1593 (1987).
	
	\bibitem{Sjoqvist2012}
	E.~Sj\"oqvist \textit{et al}
	New. J. Phys. \textbf{14}, 103035 (2012).
	
	\bibitem{Barenco1995}
	A.~Barenco \textit{et al.}
	Phys. Rev. A \textbf{52} 3457 (1995).
	
	\bibitem{Stanescu2011}
	T.~D.~Stanescu, R.~M.~Lutchyn and S.~D.~Sarma
	Phys. Rev. B \textbf{84}, 144522 (2011).
	
	\bibitem{LeijnseFlensberg2012(2)}
	M.~Leijnse and K.~Flensberg,
	Semicond. Sci. Technol. \textbf{27} 124003 (2012).
	
	\bibitem{ElliottFranz2015}
	S.~R.~Elliott and M.~Franz,
	Rev. Mod. Phys. \textbf{87}, 137 (2015).
	
	\bibitem{SatoFujimoto2016}
	M.~Sato and S.~Fujimoto
	J. Phys. Soc. Jpn. \textbf{85}, 072001 (2016).
	
	\bibitem{Ivanov2001}
	D.~A.~Ivanov,
	Phys. Rev. Lett. \textbf{86}, 268 (2001).
	
	\bibitem{FetterWalecka2003}
	A.~L.~Fetter and J.~D.~.Walecka,
	\textit{Quantum Theory of Many-Particle Systems}
	(Dover Publications, INC, Mineola, NY).

    \bibitem{MohantaTaraphder2014}
	N.~Mohanta and A.~Taraphder,
    EPL, \textbf{108}, 60001 (2014)
\end{thebibliography}
\end{document}